\documentclass[11pt]{article}

\usepackage[final]{acl}

\usepackage{times}
\usepackage{latexsym}

\usepackage[T1]{fontenc}

\usepackage[utf8]{inputenc}

\usepackage{microtype}

\usepackage{inconsolata}
\usepackage{booktabs}
\usepackage{multirow}
\usepackage{array}
\usepackage{xcolor}
\usepackage{url}
\usepackage{fvextra}
\usepackage{xspace}
\usepackage{graphicx}
\usepackage[table]{xcolor}
\usepackage{fontawesome5}

\newcommand{\benchmark}{HoosierHelp}

\DefineVerbatimEnvironment{PromptBlock}{Verbatim}{fontsize=\scriptsize,breaklines=true,breakanywhere=true}

\title{\benchmark: Benchmarking LLM Agents for Social Service Navigation}

\author{
 \textbf{Yiyang Li\textsuperscript{1}},
 \textbf{Weixiang Sun\textsuperscript{1}},
 \textbf{Tianyi Ma\textsuperscript{1}},\\
 \textbf{Kaiwen Shi\textsuperscript{1}},
 \textbf{Zheyuan Zhang\textsuperscript{1}},
 \textbf{Yanfang Ye\textsuperscript{1\textdagger}}
\\
 \textsuperscript{1}University of Notre Dame \quad \textsuperscript{\textdagger}Corresponding Author \\
 \texttt{\{yli62, yye7\}@nd.edu} \quad \faGithub\ \href{https://github.com/Yiyang-Ian-Li/HoosierHelp}{Code}
}

\begin{document}
\maketitle
\begin{abstract}
Social service navigation requires connecting help-seeking individuals to resources that satisfy their needs and specific constraints. Although LLM agents offer a promising interface for conversational resource navigation, existing benchmarks do not capture the interaction complexity and constraint-grounding demands of this setting. We introduce \textbf{\benchmark}, an interactive benchmark grounded in 3,971 Indiana public social service resources. Agents interact with simulated users, issue structured resource-search calls, handle non-ideal interactions, and select the final resources returned by the tool. \benchmark{} enhances the realism of simulated users by varying their need structure, constraint satisfiability, and behavior patterns, including impatience, rambling, unsupported requests, and self-contradiction. Experiments on 240 samples across seven LLMs show that current LLM agents remain substantially unreliable for social service navigation. Performance drops sharply on fallback-required and self-contradictory conversations, highlighting the need for agents that are more robust to complex and non-ideal user interactions.
\end{abstract}

\section{Introduction}

Access to social services is a large-scale and persistent public need. In the United States, millions of individuals and families rely on assistance to meet basic needs such as food, housing, utilities, and healthcare. In 2024, 35.9 million people lived in poverty, and 47.9 million people lived in food-insecure households~\cite{census2025income,usda2026foodsecurity}. In response to these needs, government agencies, nonprofit organizations, and community-based providers offer a wide range of social service resources, from nutrition assistance \cite{zhang2025mopi, zhang2025ngqa, huang2026glen} and housing support to healthcare access and utility relief~\cite{usda2025snap}. Meanwhile, to help people access the resources they need, social service navigation mechanisms have also been developed, including online resource directories, local service centers, and the 211 network~\footnote{\url{https://www.211.org/}}, a free information-and-referral service that connects people to community resources. Notably, in 2025, 211 network specialists made 19 million referrals for locally available help~\cite{unitedway2026impact,unitedway211about}. These numbers show that social service navigation is a critical and high-volume component of public service infrastructure.

\begin{figure}[!t]
    \centering
    \includegraphics[width=\linewidth]{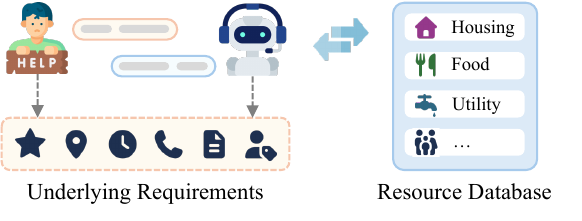}
    \caption{Illustration of social service navigation agents. The agent collects information from the user and queries the database to find the best-matched resources.}
    \label{fig:intro}
    \vspace{-10pt}
\end{figure}

\begin{figure*}[!t]
    \centering
    \includegraphics[width=\linewidth]{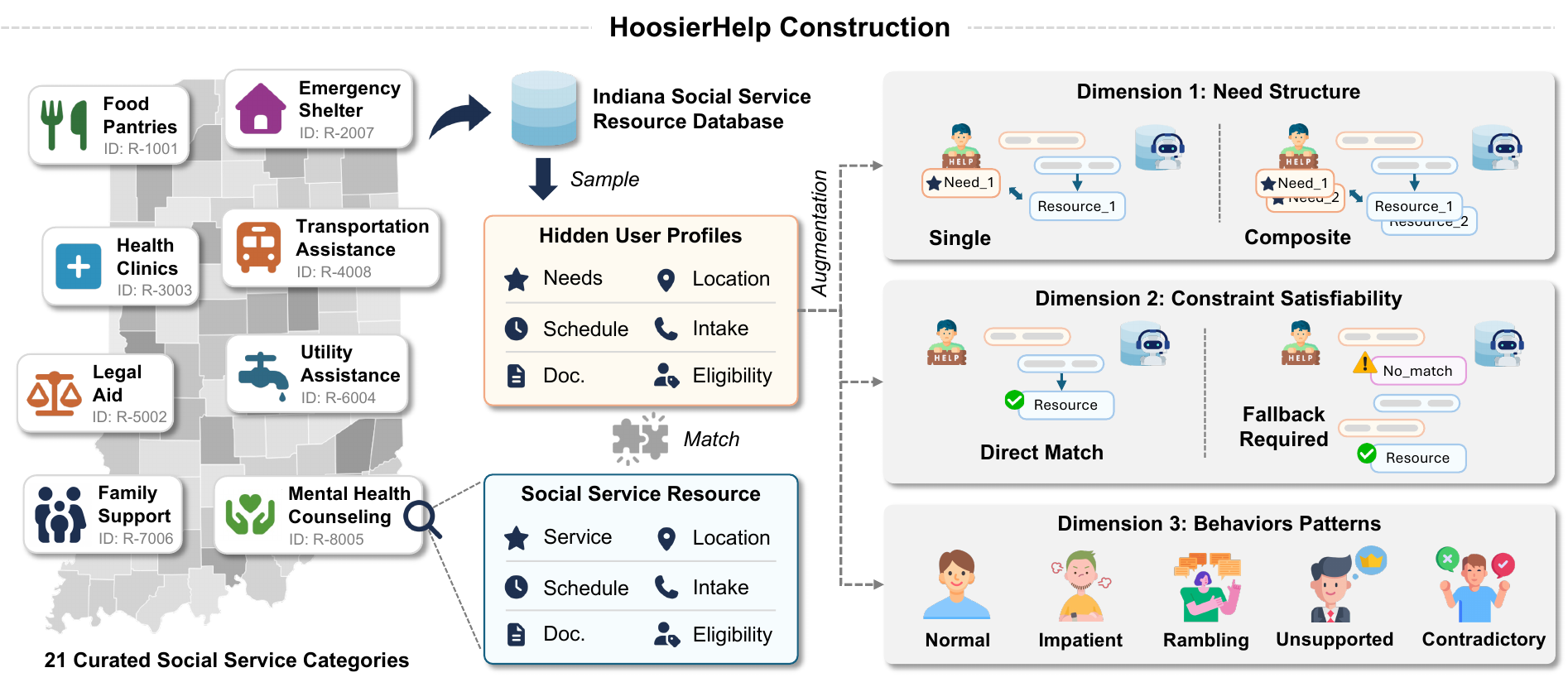}
    \caption{
    Overview of \benchmark{} construction. We build the resource database from real Indiana social services, sample user profiles with constraints associated with target resources, and augment user profiles along three dimensions: need structure, constraint satisfiability, and non-ideal user behavior.
    }
    \label{fig:construction}
    \vspace{-10pt}
\end{figure*}

However, significant challenges remain. First, providing high-quality, around-the-clock consultation and referral support is labor-intensive and costly~\cite{wa2112023expansion}. Second, help-seeking populations often include a large proportion of socially vulnerable and marginalized individuals~\cite{kreuter2012use, thompson2016promoting}, many of whom may have limited digital literacy~\cite{anuyah2023exploring, mcclain_bishop_2026_digital_divides_us}. These users face barriers when using online resource directories and often prefer conversational interaction when seeking information and support~\cite{bickmore2016improving, kocielnik2020harborbot, kocielnik2021can}.
Although general-purpose AI chatbots, e.g., ChatGPT~\cite{openai_chatgpt2026}, can help users gather information from the web, relevant resource information may be absent, and crucial details such as eligibility criteria and required documents may be missing from public webpages. These limitations motivate conversational systems that can retrieve from structured resource databases while eliciting user-specific needs and constraints.

Recent advances in LLM-based agents have created new opportunities for conversational resource navigation~\cite{schick2023toolformer, alkhouli2025confetti}. As shown in Figure~\ref{fig:intro}, in a typical resource navigation agent workflow, the agent interacts with a user to collect relevant information, invokes tools to query external databases, and then utilizes the retrieved information~\cite{wu2023inscit, lu2024stod}. 
Existing benchmarks for LLM agents primarily evaluate abilities such as API calling, multi-step tool use, and task completion in structured environments~\cite{li2023api, qin2024toolllm, yao2024tau, lu2025toolsandbox}, while failing to adequately capture the interactional complexity and domain-specific challenges of real-world social service navigation. 
Specifically, before querying the resource database, the agent needs to understand the user’s needs and constraints accurately.
However, real-world users often do not interact with agents in an idealized or fully cooperative manner~\cite{shim2025non}. Some users may make unsupported requests, give off-topic responses, or present inconsistent information. These behaviors make requirement elicitation and resource matching considerably more difficult.

To systematically study agent capabilities and behavioral patterns in this setting, we introduce \textbf{\benchmark}, an interactive benchmark for evaluating LLM agents on social service navigation with realistic simulated users. \benchmark{} is built on a curated collection of 3,971 Indiana social service resources covering 21 service categories. As illustrated in Figure~\ref{fig:construction}, \benchmark{} enhances the realism of the simulated user in three ways: (i) users may have a single need or composite needs; (ii) the user's preferred constraint may either match a resource or require fallback to an acceptable alternative; and (iii) users may exhibit various non-ideal interactive behaviors, such as impatience, rambling, unsupported request, and self-contradiction. 
Our experiments with seven state-of-the-art LLMs show that social service navigation remains challenging even for strong models. Agents often select resources without faithfully grounding the user’s full requirements, and performance degrades substantially when initially stated constraints are unsatisfiable or when users provide inconsistent information. These findings suggest that current LLM agents lack robust mechanisms for maintaining user constraints across turns and resolving non-ideal interaction patterns before making recommendations.

\begin{table*}[t]
\centering
\small
\begin{tabular}{>{\raggedright\arraybackslash}p{0.17\textwidth}>{\raggedright\arraybackslash}p{0.38\textwidth}>{\raggedright\arraybackslash}p{0.36\textwidth}}
\toprule
\textbf{User behavior} & \textbf{Definition} & \textbf{Example utterance} \\
\midrule
Normal & The user states the service need and answers the requested facts clearly and directly. & ``I am in Marion County, and Friday morning works for me.'' \\
\midrule
Impatience & The user provides available facts while sounding rushed or frustrated by repeated intake questions. & ``Yes, I can call them. Why are you asking me so many questions?'' \\
\midrule
Rambling & The user answers the question but includes distracting background or off-topic comments that should not become search facts. & ``I'm in Allen County. I keep losing track of this paperwork, and my phone has been buzzing all morning.'' \\
\midrule
Unsupported request & The user frames the real service need as something the agent cannot directly do. & ``Can you just pay my utility bill for me today? I need help with keeping the lights on.'' \\
\midrule
Self-contradictory & For one selected information area, the user gives a direct contradiction once; if asked to clarify, the user gives the real fact. & ``Thursday morning works for me, but I cannot do Thursday morning.'' \\
\bottomrule
\end{tabular}
\caption{Simulated user behaviors used in \benchmark{}, with definitions and example utterances.}
\label{tab:user-behavior-definitions}
\vspace{-8pt}
\end{table*}

\section{Benchmark Construction}

\benchmark{} evaluates social-service navigation as an interactive, tool-grounded task. In each episode, an agent must infer the user’s service need, elicit missing access constraints, search a structured resource database, and terminate with grounded resource identifiers. To construct \benchmark{}, we combine real resource data with interactive simulated users. We build a structured resource database from the public Indiana social service directory. Agents access this database through a constrained search tool interface. User messages are generated by LLM-simulated users conditioned on structured hidden profiles. We provide benchmark details and define the task in the following sections.

\subsection{Resource Database}

We build the resource database from the Indiana 211 public resource directory\footnote{\url{https://indiana211-resource.fssa.in.gov/}}. The source records include provider names, service descriptions, taxonomy labels, service areas, schedules, intake instructions, document requirements, eligibility descriptions, and other referral-relevant fields. Because many fields are stored as semi-structured or free-text descriptions, we retain only records for which the fields required for benchmark search can be parsed reliably. The resulting database contains 3,971 resources across 21 benchmark service categories. Each resource is assigned a unique \texttt{resource\_id} and represented with structured fields for \textit{service category}, \textit{location}, \textit{schedule windows}, \textit{intake methods}, \textit{document requirements}, and \textit{eligibility tags}. The 21 service categories are curated from subcategories of the source data to form a task-level taxonomy that is expressive enough for natural user requests while supporting controlled evaluation. Appendix~\ref{sec:appendix-resource-processing} gives the detailed category mapping, resource filtering counts, and tagging rules.

\subsection{Tool Design}

\benchmark{} exposes two agent-facing tools. The retrieval tool, \texttt{search\_resources}, is the only way to query the resource database. Its arguments correspond to constraints commonly used in referral matching: service category, location, schedule, intake method, available documents, and eligibility. Non-empty arguments are interpreted as hard filters across fields, while multiple values within a field are treated as alternatives. Thus, an agent should include multiple ZIP codes only when the user can use any of them, and should leave a field empty when the corresponding constraint is unrestricted.

The termination tool, \texttt{final\_recommendation}, is the only valid way to end an episode. It requires the agent to select \texttt{resource\_id} values from previously returned search results. Separating search from final recommendation also allows \benchmark{} to distinguish search errors from selection errors: an agent may formulate an appropriate query but choose the wrong returned resource, or it may choose a plausible resource after omitting or hallucinating user requirements. Appendix~\ref{sec:appendix-tool-agent} provides the detailed tool schema and mechanisms.

\subsection{User Simulation}

Social-service navigation is inherently interactive. We therefore use LLM-simulated users conditioned on structured hidden profiles. Each profile specifies the user’s service need, access constraints, and fallback options, and is paired with the target resources. 
We construct profiles with a one-to-one pairing between service needs and target resources: single-need episodes have one target resource, and composite episodes have one target resource per need, with each final satisfiable constraint state uniquely identifying its paired resource. 
The simulator observes the hidden profile and visible dialogue history, then generates user messages. This design preserves known ground truth for scoring while allowing variation in wording, order of disclosure, and conversational behavior.

\benchmark{} varies user profiles along three dimensions. First, \textbf{need structure} controls whether the user has one service need or two distinct needs. This tests whether agents can separate the user's need-specific constraints.

Second, \textbf{constraint satisfiability} controls whether the user’s initially stated constraints directly match an available resource. In direct-match cases, the stated constraints are sufficient to retrieve the target resource. In fallback-required cases, the user's initially preferred constraints yield no matching resource. The user reveals an acceptable alternative only when the agent asks about relaxing the constraint. This tests whether agents can use empty search results to revise the navigation process.

Third, \textbf{interaction behavior} controls how cooperatively the user communicates. In addition to cooperative users, \benchmark{} includes users who are impatient, rambling, make unsupported requests, or provide self-contradictory information. These modes test whether agents can extract relevant facts from noisy context, continue elicitation under pressure, recognize non-actionable requests, and resolve contradictions before search. Table~\ref{tab:user-behavior-definitions} summarizes the behavior modes.

\subsection{Task Formulation}
\label{sec:task-formulation}

Each episode is an interactive task over the resource database \(\mathcal{D}\). It is initialized by a hidden profile \(p=(n,c,f,R^\star)\), where \(n\) is the service need, \(c\) is the set of referral constraints, \(f\) contains fallback alternatives for relaxable constraints, and \(R^\star \subseteq \mathcal{D}\) is the target resource set. The agent never observes \(p\). The simulated user observes the service need, constraints, fallback alternatives, and behavior instructions; \(R^\star\) is reserved for benchmark scoring.

During the episode, the agent alternates between natural-language interaction and tool use. A call to \texttt{search\_resources} submits a structured query \(q\) over service category, location, schedule, intake method, documents, and eligibility. The benchmark executes the query against \(\mathcal{D}\) and returns all resources satisfying every non-empty field in \(q\).

The agent terminates the episode by calling \texttt{final\_recommendation} with a predicted resource set \(\hat{R}\). A prediction is valid only if every selected \texttt{resource\_id} appeared in a previous search result, ensuring that final recommendations are grounded in tool output rather than parametric knowledge. The benchmark logs the interaction trace, enabling evaluation of \(\hat{R}\) against \(R^\star\) as well as intermediate behaviors such as query formulation, fallback recovery, and grounded selection. More benchmark details are provided in Appendix~\ref{sec:appendix-benchmark-details}.

\begin{table*}[!t]
\centering
\scriptsize
\resizebox{\linewidth}{!}{
\begin{tabular}{lcc|cccc|cccc}
\toprule
\raisebox{-0.55\normalbaselineskip}{\textbf{Model}} &
\multicolumn{2}{c|}{\textbf{Overall}} &
\multicolumn{2}{c}{\textbf{Single}} &
\multicolumn{2}{c|}{\textbf{Composite}} &
\multicolumn{2}{c}{\textbf{Direct}} &
\multicolumn{2}{c}{\textbf{Fallback}} \\
\cmidrule(lr){2-3}\cmidrule(lr){4-5}\cmidrule(lr){6-7}\cmidrule(lr){8-9}\cmidrule(l){10-11}
& \textbf{Tool} & \textbf{Res.} & \textbf{Tool} & \textbf{Res.} & \textbf{Tool} & \textbf{Res.} & \textbf{Tool} & \textbf{Res.} & \textbf{Tool} & \textbf{Res.} \\
\midrule
Qwen3.7-Max & \underline{36.67} & \textbf{67.50} & \textbf{40.00} & \textbf{68.33} & \underline{33.33} & \textbf{66.67} & 49.17 & \underline{82.50} & \underline{24.17} & \underline{52.50} \\
Qwen3.6-27B & \underline{36.67} & 57.08 & \textbf{40.00} & \underline{64.17} & \underline{33.33} & 50.00 & \textbf{54.17} & 78.33 & 19.17 & 35.83 \\
Qwen3.6-35B-A3B & 23.75 & 61.67 & 25.83 & \underline{64.17} & 21.67 & 59.17 & 40.83 & \textbf{85.00} & 6.67 & 38.33 \\
GPT-5.4 & \textbf{38.75} & \underline{64.17} & 25.83 & \underline{64.17} & \textbf{51.67} & \underline{64.17} & 49.17 & 72.50 & \textbf{28.33} & \textbf{55.83} \\
GPT-OSS-120B & 34.58 & 45.42 & \underline{38.33} & 59.17 & 30.83 & 31.67 & \underline{53.33} & 60.83 & 15.83 & 30.00 \\
GPT-4.1-mini & 25.00 & 57.08 & 20.83 & 56.67 & 29.17 & 57.50 & 31.67 & 69.17 & 18.33 & 45.00 \\
DeepSeek-V4-Flash & 30.00 & 57.50 & 30.83 & 57.50 & 29.17 & 57.50 & 41.67 & 69.17 & 18.33 & 45.83 \\
\bottomrule
\end{tabular}
}
\caption{
Main results over 240 conversations across models. Tool (Tool EM) measures exact match of normalized search requirements; Res. (Resource EM) measures exact match of selected \texttt{resource\_id}s. Results are split by need structure and constraint satisfiability.
}
\label{tab:main-results}
\end{table*}

\begin{table*}[t]
\centering
\small
\definecolor{HHHeatA}{HTML}{F7FBFF}
\definecolor{HHHeatB}{HTML}{EFF6FC}
\definecolor{HHHeatC}{HTML}{E4F0F8}
\definecolor{HHHeatD}{HTML}{D9EAF5}
\definecolor{HHHeatE}{HTML}{CFE3F0}
\definecolor{HHHeatF}{HTML}{C4DDEC}
\definecolor{HHHeatG}{HTML}{B7D4E8}
\definecolor{HHHeatH}{HTML}{A9CCE3}
\definecolor{HHHeatI}{HTML}{9AC3DE}
\definecolor{HHHeatJ}{HTML}{8BB9D9}
\definecolor{HHHeatK}{HTML}{7CAFD3}
\definecolor{HHHeatL}{HTML}{6BA6CF}
\newcommand{\heatcellA}[1]{\cellcolor{HHHeatA}{#1}}
\newcommand{\heatcellB}[1]{\cellcolor{HHHeatB}{#1}}
\newcommand{\heatcellC}[1]{\cellcolor{HHHeatC}{#1}}
\newcommand{\heatcellD}[1]{\cellcolor{HHHeatD}{#1}}
\newcommand{\heatcellE}[1]{\cellcolor{HHHeatE}{#1}}
\newcommand{\heatcellF}[1]{\cellcolor{HHHeatF}{#1}}
\newcommand{\heatcellG}[1]{\cellcolor{HHHeatG}{#1}}
\newcommand{\heatcellH}[1]{\cellcolor{HHHeatH}{#1}}
\newcommand{\heatcellI}[1]{\cellcolor{HHHeatI}{#1}}
\newcommand{\heatcellJ}[1]{\cellcolor{HHHeatJ}{#1}}
\newcommand{\heatcellK}[1]{\cellcolor{HHHeatK}{#1}}
\newcommand{\heatcellL}[1]{\cellcolor{HHHeatL}{#1}}
\resizebox{\linewidth}{!}{
\begin{tabular}{lccccc|ccccc}
\toprule
\raisebox{-0.55\normalbaselineskip}{\textbf{Model}} &
\multicolumn{5}{c|}{\textbf{Tool EM}} &
\multicolumn{5}{c}{\textbf{Resource EM}} \\
\cmidrule(lr){2-6}\cmidrule(l){7-11}
& \textbf{Normal} & \textbf{Impat.} & \textbf{Ramble} & \textbf{Unsup.} & \textbf{Contra.}
& \textbf{Normal} & \textbf{Impat.} & \textbf{Ramble} & \textbf{Unsup.} & \textbf{Contra.} \\
\midrule
Qwen3.7-Max & \heatcellI{66.67} & \heatcellF{43.75} & \heatcellE{33.33} & \heatcellE{31.25} & \heatcellB{8.33} & \heatcellL{85.42} & \heatcellK{79.17} & \heatcellI{60.42} & \heatcellJ{70.83} & \heatcellF{41.67} \\
Qwen3.6-27B & \heatcellI{64.58} & \heatcellE{33.33} & \heatcellF{43.75} & \heatcellE{35.42} & \heatcellA{6.25} & \heatcellJ{72.92} & \heatcellK{77.08} & \heatcellG{52.08} & \heatcellJ{70.83} & \heatcellB{12.50} \\
Qwen3.6-35B-A3B & \heatcellF{43.75} & \heatcellD{22.92} & \heatcellC{16.67} & \heatcellC{20.83} & \heatcellB{14.58} & \heatcellI{64.58} & \heatcellJ{70.83} & \heatcellI{60.42} & \heatcellI{62.50} & \heatcellG{50.00} \\
GPT-5.4 & \heatcellH{58.33} & \heatcellG{52.08} & \heatcellF{41.67} & \heatcellE{33.33} & \heatcellB{8.33} & \heatcellK{81.25} & \heatcellK{81.25} & \heatcellI{66.67} & \heatcellJ{72.92} & \heatcellC{18.75} \\
GPT-OSS-120B & \heatcellG{52.08} & \heatcellG{47.92} & \heatcellF{37.50} & \heatcellD{29.17} & \heatcellA{6.25} & \heatcellH{56.25} & \heatcellH{56.25} & \heatcellF{43.75} & \heatcellG{47.92} & \heatcellD{22.92} \\
GPT-4.1-mini & \heatcellG{47.92} & \heatcellF{41.67} & \heatcellB{14.58} & \heatcellB{14.58} & \heatcellA{6.25} & \heatcellL{83.33} & \heatcellJ{70.83} & \heatcellI{62.50} & \heatcellH{58.33} & \heatcellB{10.42} \\
DeepSeek-V4-Flash & \heatcellH{54.17} & \heatcellG{50.00} & \heatcellD{25.00} & \heatcellB{10.42} & \heatcellB{10.42} & \heatcellK{81.25} & \heatcellJ{72.92} & \heatcellH{58.33} & \heatcellG{50.00} & \heatcellD{25.00} \\
\bottomrule
\end{tabular}
}
\caption{
Performance by simulated-user behavior. Tool EM and Resource EM are exact-match rates for search requirements and selected resources, respectively. Impat., Unsup., and Contra. denote impatient, unsupported-request, and self-contradictory behaviors; darker cells indicate higher scores.
}
\vspace{-10pt}
\label{tab:user-behavior-results}
\end{table*}

\begin{table*}[t]
\centering
\small
\definecolor{HHTurnA}{HTML}{F7FBFF}
\definecolor{HHTurnB}{HTML}{EFF6FC}
\definecolor{HHTurnC}{HTML}{E4F0F8}
\definecolor{HHTurnD}{HTML}{D9EAF5}
\definecolor{HHTurnE}{HTML}{CFE3F0}
\definecolor{HHTurnF}{HTML}{C4DDEC}
\definecolor{HHTurnG}{HTML}{B7D4E8}
\definecolor{HHTurnH}{HTML}{A9CCE3}
\definecolor{HHTurnI}{HTML}{9AC3DE}
\definecolor{HHTurnJ}{HTML}{8BB9D9}
\definecolor{HHTurnK}{HTML}{7CAFD3}
\definecolor{HHTurnL}{HTML}{6BA6CF}
\newcommand{\turncellA}[1]{\cellcolor{HHTurnA}{#1}}
\newcommand{\turncellB}[1]{\cellcolor{HHTurnB}{#1}}
\newcommand{\turncellC}[1]{\cellcolor{HHTurnC}{#1}}
\newcommand{\turncellD}[1]{\cellcolor{HHTurnD}{#1}}
\newcommand{\turncellE}[1]{\cellcolor{HHTurnE}{#1}}
\newcommand{\turncellF}[1]{\cellcolor{HHTurnF}{#1}}
\newcommand{\turncellG}[1]{\cellcolor{HHTurnG}{#1}}
\newcommand{\turncellH}[1]{\cellcolor{HHTurnH}{#1}}
\newcommand{\turncellI}[1]{\cellcolor{HHTurnI}{#1}}
\newcommand{\turncellJ}[1]{\cellcolor{HHTurnJ}{#1}}
\newcommand{\turncellK}[1]{\cellcolor{HHTurnK}{#1}}
\newcommand{\turncellL}[1]{\cellcolor{HHTurnL}{#1}}
\resizebox{\linewidth}{!}{
\begin{tabular}{lc|cc|cc|ccccc}
\toprule
\raisebox{-0.55\normalbaselineskip}{\textbf{Model}} &
\multicolumn{1}{c|}{\textbf{Overall}} &
\multicolumn{2}{c|}{\textbf{Task Type}} &
\multicolumn{2}{c|}{\textbf{Constraint}} &
\multicolumn{5}{c}{\textbf{User Behavior}} \\
\cmidrule(lr){2-2}\cmidrule(lr){3-4}\cmidrule(lr){5-6}\cmidrule(l){7-11}
& \textbf{All} & \textbf{Single} & \textbf{Comp.} & \textbf{Direct} & \textbf{Fallback}
& \textbf{Normal} & \textbf{Impat.} & \textbf{Ramble} & \textbf{Unsup.} & \textbf{Contra.} \\
\midrule
Qwen3.7-Max & \turncellH{5.98} & \turncellH{6.12} & \turncellG{5.84} & \turncellF{5.18} & \turncellJ{6.78} & \turncellG{5.81} & \turncellD{4.58} & \turncellI{6.40} & \turncellH{5.92} & \turncellK{7.19} \\
Qwen3.6-27B & \turncellG{5.85} & \turncellG{5.67} & \turncellH{6.03} & \turncellE{4.99} & \turncellI{6.70} & \turncellG{5.88} & \turncellB{3.73} & \turncellI{6.46} & \turncellG{5.52} & \turncellL{7.65} \\
Qwen3.6-35B-A3B & \turncellC{4.00} & \turncellC{3.90} & \turncellC{4.10} & \turncellA{3.34} & \turncellD{4.66} & \turncellD{4.33} & \turncellB{3.69} & \turncellC{3.88} & \turncellC{4.02} & \turncellC{4.08} \\
GPT-5.4 & \turncellH{6.28} & \turncellI{6.47} & \turncellH{6.08} & \turncellG{5.53} & \turncellJ{7.02} & \turncellH{6.23} & \turncellG{5.58} & \turncellH{6.12} & \turncellG{5.71} & \turncellL{7.73} \\
GPT-OSS-120B & \turncellI{6.65} & \turncellH{5.94} & \turncellK{7.36} & \turncellH{5.99} & \turncellK{7.31} & \turncellI{6.42} & \turncellH{6.06} & \turncellI{6.73} & \turncellI{6.56} & \turncellK{7.48} \\
GPT-4.1-mini & \turncellG{5.88} & \turncellH{6.07} & \turncellG{5.68} & \turncellE{5.07} & \turncellI{6.68} & \turncellG{5.81} & \turncellF{5.17} & \turncellG{5.58} & \turncellF{5.27} & \turncellK{7.54} \\
DeepSeek-V4-Flash & \turncellI{6.50} & \turncellI{6.58} & \turncellI{6.42} & \turncellG{5.75} & \turncellK{7.25} & \turncellH{6.27} & \turncellF{5.42} & \turncellI{6.60} & \turncellI{6.50} & \turncellL{7.71} \\
\bottomrule
\end{tabular}
}
\caption{
Average interaction turns by benchmark setting. Comp., Impat., Unsup., and Contra. denote composite, impatient, unsupported-request, and self-contradictory settings; darker cells indicate longer conversations.
}
\vspace{-10pt}
\label{tab:turn-results}
\end{table*}

\section{Experiments}

\subsection{Experimental Setup and Metrics}

We evaluate seven LLMs: Qwen3.7-Max~\cite{qwen37}, Qwen3.6-27B~\cite{qwen3.6-27b}, Qwen3.6-35B-A3B~\cite{qwen36_35b_a3b}, GPT-5.4~\cite{openai2025gpt54}, GPT-OSS-120B~\cite{openai2025gptoss}, GPT-4.1-mini~\cite{openai2025gpt41}, and DeepSeek-V4-Flash~\cite{deepseekai2026deepseekv4}. All runs use the same 240 samples with a maximum of eight interaction turns and use GPT-4.1-mini as the simulated-user model. The evaluation set is balanced across the benchmark dimensions, so each reported split compares equally sized subsets. 

We report two primary outcome metrics. \textbf{Tool EM} is exact match over \texttt{search\_resources} arguments. It measures whether the agent preserved the full structured constraints. \textbf{Resource EM} is exact match over the resource IDs selected in the \texttt{final\_recommendation} tool call. More details are provided in Appendix~\ref{sec:appendix-experiment}.

\subsection{Tool Grounding and Referral Accuracy}

Table~\ref{tab:main-results} shows a consistent gap between tool grounding and final referral accuracy. Resource EM is substantially higher than Tool EM for every model. The best Resource EM is 67.50\% for Qwen3.7-Max, while the best Tool EM is 38.75\% for GPT-5.4. This gap is central to the benchmark: agents often reach plausible or correct resources without faithfully preserving the full structured requirement state. In social-service referral, such behavior is risky because omitted constraints can affect eligibility, document requirements, intake options, or service availability.

The models exhibit different tradeoffs. Qwen3.7-Max is the strongest resource selector overall and remains relatively stable across single and composite cases, but its Tool EM stays below 40\%. GPT-5.4 achieves the strongest Tool EM, especially on composite cases, suggesting better preservation of need-specific constraints. Qwen3.6-35B-A3B achieves competitive Resource EM despite much lower Tool EM, indicating a more aggressive recommendation style that often selects plausible resources before fully grounding the search.

The need-structure split is less diagnostic than the constraint-satisfiability split. Composite conversations are not uniformly harder: GPT-5.4, GPT-4.1-mini, and DeepSeek-V4-Flash have similar or higher Tool EM on composite cases than on single-need cases. This suggests that the primary challenge in \benchmark{} is not simply tracking multiple needs, but handling constraints that are missing, noisy, contradictory, or initially unsatisfiable.

\begin{figure}[!t]
    \centering
    \includegraphics[width=\linewidth]{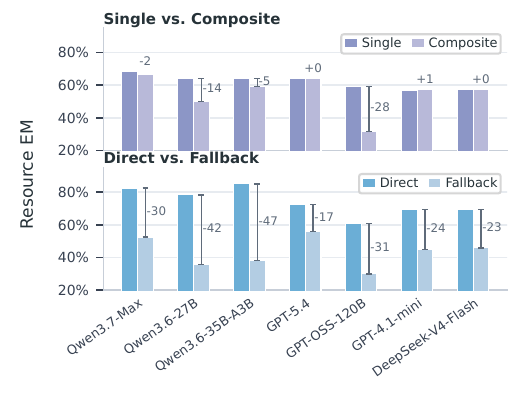}
    \caption{Resource EM by need structure and constraint satisfiability. Numbers above bars show absolute differences between paired settings.}
    \label{fig:gap}
    \vspace{-10pt}
\end{figure}

\subsection{Fallback Reasoning and User Behavior}

The constraint-satisfiability (i.e., direct/fallback) split in Table~\ref{tab:main-results} and Figure~\ref{fig:gap} is the clearest stress test. Every model drops substantially on fallback-required cases. For example, Qwen3.6-35B-A3B falls from 85.00\% Resource EM on direct-match cases to 38.33\% on fallback-required cases. GPT-5.4 is the most robust on fallback cases, but still drops from 72.50\% to 55.83\% Resource EM.

Fallback-required cases require agents to use tool feedback as part of the dialogue state. A successful agent must search with the user’s stated constraint, observe that no resource matches, ask whether the relevant constraint can be relaxed, and then revise only that constraint while preserving all other facts. The large fallback drop shows that many agents either treat empty results as an endpoint, broaden the search too aggressively, or lose previously elicited constraints when retrying.

Table~\ref{tab:user-behavior-results} further shows that user behavior changes both accuracy and failure mode. Normal conversations are not solved: the best normal Resource EM is 85.42\%, and the best normal Tool EM is 66.67\%. Self-contradictory users are the hardest condition. Tool EM falls below 15\% for every model, and Resource EM often collapses as well, indicating that agents rarely recognize contradictions as unresolved facts that require clarification before search. Rambling and unsupported-request users stress a different capability: separating actionable referral constraints from irrelevant context or requests the tool cannot satisfy. Several models retain moderate Resource EM while losing Tool EM, suggesting that they recover enough signal to select plausible resources but fail to maintain a faithful structured account of the user’s requirements. Appendix~\ref{sec:appendix-combined-breakdown} shows that these patterns persist when splitting by the full combination of need structure, constraint satisfiability, and user behavior.

\begin{figure}[t]
    \centering
    \includegraphics[width=\linewidth]{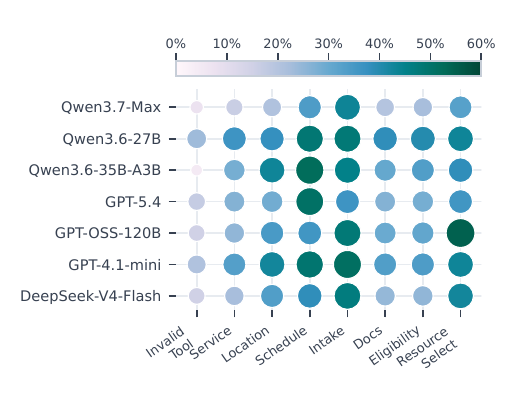}
    \caption{Field-level failure rates across models. Larger, darker points indicate more frequent failures.}
    \label{fig:failure}
    \vspace{-10pt}
\end{figure}

\subsection{Interaction Length and Diagnostic Errors}

Table~\ref{tab:turn-results} shows that interaction length reflects agent strategy rather than quality alone. Qwen3.6-35B-A3B averages only 4.00 agent turns and frequently reaches a final recommendation, but its Tool EM is low. In contrast, GPT-5.4 averages 6.28 turns and obtains the highest Tool EM. This contrast suggests two broad styles: fast recommendation-oriented behavior that often finds plausible resources with incomplete grounding, and slower elicitation-oriented behavior that better preserves structured requirements. The longest conversations occur in fallback-required and self-contradictory settings, but additional turns do not reliably close the performance gap. Extra interaction helps only when agents use it to ask the right clarification question, update the correct constraint, and preserve previously collected information. Thus, short conversations are not necessarily efficient, and long conversations are not necessarily robust.

Figure~\ref{fig:failure} provides a field-level view of where agents fail. Invalid tool use is relatively rare, so most errors are not caused by inability to invoke the tool. Instead, failures concentrate in requirement fields and final selection. Schedule and intake errors are common across models, reflecting the difficulty of mapping conversational availability and access preferences into structured filters. Location and service-category errors also appear frequently, especially for models with lower Tool EM. Document and eligibility errors are less uniformly dominant but remain important because they affect whether a referral is practically usable. The final resource-selection column shows that even after a tool call succeeds syntactically, agents may still select the wrong returned resource or fail to provide a grounded final recommendation. 

Overall, the experiments show that current LLM agents remain unreliable for social-service navigation. The main failure is not simply retrieving a relevant resource, but maintaining a faithful requirement state through noisy interaction, tool feedback, fallback revision, and grounded final selection.

\section{Conclusion}

We introduced \benchmark{}, an interactive benchmark for evaluating LLM agents on social service navigation. \benchmark{} tests whether agents can elicit user constraints, issue structured search calls, recover from fallback constraints, and make grounded final recommendations under realistic simulated user behaviors. Experiments across seven LLMs show that current agents remain unreliable: they often select plausible resources without faithfully grounding user requirements, and they degrade sharply when constraints are initially unsatisfiable or self-contradictory. These findings highlight the need for agents with stronger information elicitation and grounded tool use.

\section*{Acknowledgements}
This work was partially supported by the NSF under grants IIS-2533550, IIS-2321504, IIS-2217239, CNS-2426514, and CMMI-2146076, Notre Dame Strategic Framework Research Grant (2025), and Notre Dame Poverty Research Package (2025). Any expressed opinions, findings, and conclusions or recommendations are those of the authors and do not necessarily reflect the views of the sponsors.

\section*{Limitations}

\benchmark{} is a benchmark for evaluating referral agents, not a live referral service. The resource index is filtered and normalized for experimental use and may not reflect current availability, eligibility, or provider capacity. The ground-truth construction intentionally enforces unique answers, which supports automatic scoring but simplifies the reality that multiple resources may be reasonable for a person in need. The user simulator is also an approximation: its behavior modes are designed to stress agent behavior, but they cannot capture the full range of lived experience, crisis communication, language access needs, or trust dynamics involved in real social-service navigation. Any deployment of agents in this domain should include human oversight, escalation paths, and ongoing validation against current provider information.

\section*{Ethical Considerations}

Social service referral is a high-stakes domain involving people who may face poverty, housing instability, food insecurity, health needs, or other vulnerabilities. \benchmark{} is intended only for evaluation and should not be treated as a live referral system. Its resources are filtered for experimental use and may not reflect current availability, eligibility rules, or intake procedures. Any deployment of referral agents should include human oversight, escalation paths, provider validation, and crisis-handling mechanisms.

\benchmark{} uses simulated users rather than real help-seeking individuals, reducing privacy risks and avoiding exposure of vulnerable users to experimental systems. However, simulation cannot capture the full range of lived experiences, communication styles, language needs, disability access needs, or trust dynamics in social-service navigation. We therefore frame \benchmark{} as a tool for identifying agent limitations rather than certifying deployment readiness, and encourage participatory evaluation and monitoring for disparate impacts before real-world use.

\bibliography{custom}

\appendix

\section{Related Work}

\subsection{Tool-Calling Agent Evaluation}

LLMs have advanced rapidly in recent years \cite{ma2025llm, ye2025llms4all, shi2026sage, ma2026non}. Building on this progress, LLM-driven agents have gained prominence for their ability to plan, interact, and solve complex tasks with limited human oversight \cite{zhang2026mapro, ma2025autodata, chen2025obvious,chen2025clear,zhang2025agentrouter, shi2026ng, huang2026evolverouter, bao2026drift, li2026longda}. Building on this trend, recent work increasingly evaluates large language models as agents that can interact with external tools. Existing benchmarks study whether agents can select APIs, construct valid function calls, compose multiple tool invocations, and complete tasks in executable environments~\cite{li2023api, qin2024toolllm, patil2024gorilla}. More recent benchmarks extend this evaluation to interactive settings, where agents must converse with simulated users, follow domain policies, maintain state, and decide when to call tools~\cite{alkhouli2025confetti, yao2024tau, lu2025toolsandbox}. While these benchmarks have advanced the evaluation of tool-calling agents, they are not designed for social service referral, where agents must elicit missing requirements, ground partially specified needs into structured search parameters, and match users to resources under specific requirements and constraints.

\subsection{Conversational Social Service Navigation}

Social service access often depends on information-and-referral processes that connect people to resources matching their needs and circumstances. Prior work suggests that web-based resource search can be difficult for users facing digital literacy, health literacy, language, or socioeconomic barriers, and that conversational interfaces can make information seeking more understandable, personal, and supportive~\cite{bickmore2016improving, kocielnik2020harborbot, kocielnik2021can}. AI-driven methods \cite {zhao2023self,ju2023graphpatcher,ju2022grape,qian2022co,zhao2021multi,chen2018droideye,ye2009intelligent} have been used to tackle these challenges. Furthermore, recent work has explored chatbots for social-needs screening and resource sharing, demonstrating the promise of conversational approaches in healthcare and community settings~\cite{sezgin2024chatbot}. However, this work primarily studies system design, user experience, and feasibility rather than systematic benchmark evaluation. \benchmark{} instead evaluates LLM agents for realistic, interaction-sensitive social service referral.

\section{Additional Benchmark Details}
\label{sec:appendix-benchmark-details}

\subsection{Resource Processing}
\label{sec:appendix-resource-processing}

Table~\ref{tab:appendix-filtering} summarizes the resource filtering pipeline. The filter is intended to make the search tool executable and automatically scorable, not to claim that excluded records are unimportant in real referral practice. In particular, statewide resources and records with ambiguous schedules or intake instructions are excluded because the benchmark focuses on grounding local, structured constraints.

\begin{table}[h]
\centering
\small
\begin{tabular}{lrr}
\toprule
\textbf{Step} & \textbf{Kept} & \textbf{Removed} \\
\midrule
Raw deduplicated records & 9,987 & -- \\
In-state local service area & 8,787 & 1,200 \\
Parseable schedule & 7,423 & 1,364 \\
Parseable or empty documents & 4,863 & 2,560 \\
Parseable intake methods & 4,661 & 202 \\
Benchmark metadata & 3,971 & 690 \\
\bottomrule
\end{tabular}
\caption{Filtering pipeline for the Indiana social service resource database.}
\label{tab:appendix-filtering}
\end{table}

Table~\ref{tab:appendix-category-map} shows how raw Indiana 211 subcategories are collapsed into the 21 benchmark service categories. Several raw administrative or informational subcategories are excluded because they do not correspond to actionable local referral needs in the current benchmark.

\begin{table*}[t]
\centering
\scriptsize
\resizebox{\linewidth}{!}{
\begin{tabular}{p{0.24\textwidth}p{0.70\textwidth}}
\toprule
\textbf{Benchmark category} & \textbf{Mapped raw Indiana 211 subcategories} \\
\midrule
Community and Recreation & Arts and Culture; Community Facilities/Centers; Leisure Activities/Recreation \\
Disability and Rehabilitation & Rehabilitation/Habilitation Services \\
Disaster and Environmental Services & Disaster Services; Environmental Protection and Improvement \\
Education and Youth Programs & Educational Institutions/Schools; Educational Programs; Educational Support Services; Social Development and Enrichment \\
Employment and Job Training & Employment \\
Family and Caregiver Services & Individual and Family Support Services \\
Financial Assistance and Benefits & Money Management; Public Assistance Programs; Social Insurance Programs; Temporary Financial Assistance \\
Food Assistance & Food \\
Health Care & Emergency Medical Care; Health Screening/Diagnostic Services; Inpatient Health Facilities; Outpatient Health Facilities; Public Health; Specialty Medicine \\
Housing and Shelter & Housing/Shelter \\
Legal and Court Help & Courts; Judicial Services; Legal Services \\
Material Goods & Material Goods \\
Medical Support Services & Health Supportive Services; Specialized Treatment and Prevention \\
Mental Health Care & Mental Health Assessment and Treatment; Mental Health Care Facilities; Mental Health Support Services; Mutual Support \\
Pet and Animal Services & Domestic Animal Services \\
Pregnancy and Reproductive Health & Human Reproduction \\
Public Safety & Law Enforcement Agencies; Law Enforcement Services; Public Safety \\
Substance Use Services & Substance Use Disorder Services \\
Tax Help & Tax Organizations and Services \\
Transportation & Transportation \\
Utility Assistance & Utilities \\
\bottomrule
\end{tabular}
}
\caption{Mapping from raw Indiana 211 subcategories to the 21 benchmark service categories. Dropped raw subcategories include broad administrative, consumer-regulation, correctional, information-only, military-service, and civic-participation categories that are not used as target service needs.}
\label{tab:appendix-category-map}
\end{table*}

Table~\ref{tab:appendix-tagging} summarizes the resource tagging rules used after category mapping. These rules are intentionally conservative: resources with ambiguous core fields are filtered out rather than forced into unsupported structured values.

\begin{table*}[t]
\centering
\small
\begin{tabular}{p{0.18\textwidth}p{0.26\textwidth}p{0.47\textwidth}}
\toprule
\textbf{Field} & \textbf{Source text} & \textbf{Tagging rule} \\
\midrule
Schedule windows & \texttt{site\_schedule} & Parse structured days and time intervals into windows over \texttt{mon}--\texttt{sun}. Resources are retained only when the schedule is parseable and non-empty. 24-hour windows are represented separately. \\
\midrule
Service categories & Raw subcategories & Map raw subcategories through Table~\ref{tab:appendix-category-map}; keep resources with at least one and at most three benchmark service categories. \\
\midrule
Intake methods & \texttt{site\_details} & Detect evidence for \texttt{call}, \texttt{walk\_in}, \texttt{online}, \texttt{appointment}, \texttt{email}, \texttt{text}, and \texttt{mail}. Resources without parseable intake evidence are removed. \\
\midrule
Document requirements & \texttt{documents\_required} & Tag supported documents: \texttt{photo\_id}, \texttt{proof\_of\_income}, \texttt{proof\_of\_address}, \texttt{lease}, \texttt{insurance\_card}, \texttt{social\_security}, \texttt{birth\_certificate}, and \texttt{utility\_bill}. Explicit no-document cases are tagged as \texttt{none}; ambiguous cases such as ``call for details'' are filtered. \\
\midrule
Eligibility & \texttt{site\_eligibility} & Tag common eligibility concepts: \texttt{low\_income}, \texttt{resident}, \texttt{homeless}, \texttt{veteran}, \texttt{senior}, \texttt{youth}, \texttt{family}, \texttt{pregnant}, \texttt{disability}, \texttt{uninsured}, and \texttt{medicaid}; use \texttt{none} when no supported tag is found. \\
\bottomrule
\end{tabular}
\caption{Structured tagging rules for benchmark resources. The same tags define the agent-facing tool schema and the hidden user constraints used for scoring.}
\label{tab:appendix-tagging}
\end{table*}

\subsection{Tool Schema}
\label{sec:appendix-tool-agent}

The agent-facing tool interface contains \texttt{search\_resources} and \texttt{final\_recommendation}. Table~\ref{tab:appendix-search-tool} gives the full argument semantics for \texttt{search\_resources}. The tool uses conjunctive semantics across non-empty fields: service, schedule, location, intake, documents, and eligibility must all match. Values within one field are disjunctive alternatives. The three location fields are also alternatives across county, city, and ZIP code, so an agent should include every location the user says they can accept.

\begin{table*}[t]
\centering
\small
\begin{tabular}{p{0.20\textwidth}p{0.22\textwidth}p{0.50\textwidth}}
\toprule
\textbf{Argument} & \textbf{Values} & \textbf{Semantics} \\
\midrule
\texttt{service\_categories} & List over the 21 benchmark categories & Required. A resource matches when any requested service category appears in its tagged service categories. \\
\midrule
\texttt{counties} & Uppercase county names, e.g., \texttt{MARION} & Optional location alternatives. If any location field is non-empty, a resource must match at least one requested county, city, or ZIP code. \\
\texttt{cities} & City names, normalized for matching & Optional location alternatives. Agents should not infer a county from a city unless the user stated it. \\
\texttt{zipcodes} & ZIP-code strings & Optional location alternatives. \\
\midrule
\texttt{schedule.day} & \texttt{mon}, \texttt{tue}, \texttt{wed}, \texttt{thu}, \texttt{fri}, \texttt{sat}, \texttt{sun} & Optional. Use only when the user gives a concrete day. Day alone means the user can use any open time on that day. \\
\texttt{schedule.time} & 24-hour \texttt{HH:MM} & Optional. Use when the user gives a specific time on a concrete day. \\
\texttt{schedule.start\_time}, \texttt{schedule.end\_time} & 24-hour \texttt{HH:MM} & Optional. Use for an availability window on a concrete day. \\
\texttt{schedule.requires\_24} & Boolean & Optional. Use only when the user requires 24-hour availability. \\
\midrule
\texttt{intake\_methods} & \texttt{call}, \texttt{walk\_in}, \texttt{online}, \texttt{appointment}, \texttt{email}, \texttt{text}, \texttt{mail} & Optional hard filter over access methods. If the user has no preference, leave empty rather than listing all methods. \\
\midrule
\texttt{available\_documents} & Supported document tags in Table~\ref{tab:appendix-tagging} & Optional. Represents documents the user can provide. A resource matches when its required concrete documents are a subset of the user's available documents. \\
\midrule
\texttt{eligibility} & Supported eligibility tags in Table~\ref{tab:appendix-tagging} & Optional. Represents user attributes relevant to eligibility. A resource with concrete eligibility requirements must be satisfied by the supplied tags. \\
\bottomrule
\end{tabular}
\caption{Argument semantics for \texttt{search\_resources}. Empty optional fields mean no user constraint; they are not interpreted as all possible values.}
\label{tab:appendix-search-tool}
\end{table*}

The tool returns a JSON object with a \texttt{resources} list. Each returned item contains \texttt{resource\_id}, \texttt{resource\_name}, city, ZIP code, service categories, intake methods, concrete document requirements, and eligibility tags. Results are ranked by matched-field score and then deterministically by resource name and ID, with a protocol default limit of 10. \texttt{final\_recommendation} takes a list of selected \texttt{resource\_id} values and a short message, and is the only valid final action. Resource EM is computed over the IDs supplied to this final tool.

\subsection{User Simulation}
\label{sec:appendix-user-simulation}

The simulated user is an LLM conditioned on a structured user profile and the visible conversation state. The profile includes case type, constraint profile, service needs, ordinary constraints, and first-choice versus fallback constraints when applicable. The simulator is instructed to reveal only facts that the agent asks for, to avoid mentioning resource IDs or benchmark field names, and to treat fallback options as unavailable until an empty search result has occurred and the agent asks whether location, schedule, or intake can change. Self-contradictory behavior selects one concrete information area and produces a single direct contradiction before resolving it if the agent asks for clarification.

\subsection{Prompts}
\label{sec:appendix-prompts}

For reproducibility, we include the core prompts used by the agent and simulated user. The agent prompt below is paired with the tool schema in Appendix~\ref{sec:appendix-tool-agent}; for API backends with native function calling, the Qwen-style XML examples are omitted while the protocol instructions are unchanged.

\begin{PromptBlock}
You are an Indiana 211 resource-search agent.

Follow this protocol:

1. Understand the need.
- Identify whether the user has one need or two different needs.
- If the user has only named a broad service need, ask a clarification question before searching.
- Do not invent facts or constraints.

2. Collect search facts before searching.
For each need, learn enough to fill the search_resources arguments:
- service category
- location
- schedule
- intake method
- available documents
- eligibility

For each field, either collect the user's constraint or establish that the user
has no constraint for that field. If the user has no requirement for a field,
leave that field empty. If the user gives multiple acceptable locations, times,
or intake methods, include every acceptable value in the tool arguments.
Treat "no preference", "any time", "any method", "no restrictions", and similar
answers as no constraint: leave the corresponding schedule or list field empty
rather than enumerating every possible value. Do not infer a county from a city
or ZIP; include a county only when the user explicitly gives a county.
Use a schedule object only when the user gives a concrete day or time window;
do not represent open availability as Monday all day or any other default day.
When the user changes location for a fallback search, use the new accepted
location instead of combining it with the original city, county, or ZIP, unless
the user explicitly says both locations should remain acceptable.
Do not search after learning only the service need and location. Before the
first search, ask about schedule, intake method, available documents, and
eligibility unless the user has already provided those facts or clearly said
they have no constraints for them.
Treat an early search with only service category and location as an error
unless the user has already said there are no other constraints.

3. Keep questions low-burden.
- Do not ask for location, schedule, intake method, documents, and eligibility all in one message.
- Ask the next useful small group of facts based on what is still missing and what the user has already provided.
- Example flow: after the service need is clear, ask location first, then ask schedule plus intake method, then ask documents plus eligibility. This is an example, not a required script.
- If the user does not answer one requested field, ask for it once more only if it is needed to avoid guessing. Do not keep asking the same clarification after the user gives a usable answer or says they have no constraint.

4. Handle two-need conversations naturally.
- Search for each need with a separate search_resources call unless the needs truly have identical constraints and service categories.
- Confirm shared facts once when useful.
- Ask separately for facts that may differ by need, especially schedule and intake method.
- Return one selected resource_id for each need.

5. Search and handle results.
- Use only values allowed by the tool schema.
- When you are ready to search, call search_resources in that same assistant turn. Do not only say that you will search.
- After tool results are provided, do not repeat the same search unless the user clearly provided new search constraints.
- If a search returns one or more resources for the current user constraints, the next assistant action should be final_recommendation.
- If a search returns no resources, ask the user whether location, schedule, or intake constraints can be adjusted. 
- If the user gives a fallback value, preserve all previously collected facts in the next search. If some ordinary facts were never collected, ask for them before making the fallback search.

6. End only with final_recommendation.
- After matching tool results are available, end by calling the final_recommendation tool with the selected resource_id or resource_ids.
- When you are ready to recommend resources, call final_recommendation in that same assistant turn. Do not only say that you will make a recommendation.
- A normal sentence is not a valid final answer.

For Qwen-style local tool calls, emit <tool_call> blocks containing JSON objects
with "name" and "arguments" keys, never "parameters":
<tool_call>
{"name": "search_resources", "arguments": {"service_categories": ["..."], "schedule": {}, "counties": [], "cities": [], "zipcodes": [], "intake_methods": [], "available_documents": [], "eligibility": []}}
</tool_call>

For final_recommendation, put selected resource_id values in resource_ids
exactly as returned by search_resources:
<tool_call>
{"name": "final_recommendation", "arguments": {"resource_ids": ["in211-..."], "message": "I recommend in211-... because it matches the user's need and constraints."}}
</tool_call>
\end{PromptBlock}

The simulated-user system prompt is:

\begin{PromptBlock}
You are simulating a person asking for help finding Indiana community resources.

You must act like the user, not like an assistant. Do not mention that you are
simulated, following a behavior pattern, using hidden facts, or obeying
instructions.

Use the user profile as what this person knows about their own situation.
Answer the agent's latest question naturally and do not volunteer search facts
the agent did not ask for. Follow the behavior instructions when they say to be
self-contradictory, impatient, tangential, or unrealistic.

Keep the response as a natural user message, not an intake form or organized
checklist. When giving a time range, make AM/PM or 24-hour meaning clear for
both the start and end time if known. Do not use lists unless the user would
naturally list a few items.
\end{PromptBlock}

Each simulated-user turn is generated from the behavior instruction, turn type,
conversation state, and structured user profile:

\begin{PromptBlock}
[Behavior instruction for the selected user behavior]

Current turn type: opening or follow-up.

[Turn-specific instruction]

User profile:
{
  "case_type": "single" or "composite",
  "constraint_profile": "direct_match" or "fallback_required",
  "needs": [
    {
      "plain_language_need": "...",
      "service_categories": ["..."],
      "documents": [...],
      "eligibility": [...],
      "constraints": {
        "location": {...},
        "schedule": {...},
        "intake_methods": [...]
      }
    }
  ],
  "fallback_rules": {
    "first_choice_stage": "Before an empty search result, use first_choice for location, schedule, and intake constraints when asked.",
    "fallback_stage": "After an empty search result, fallback can be mentioned only if the agent asks whether location, schedule, or intake can change.",
    "changeable_fields": ["location", "schedule", "intake"],
    "not_changeable_fields": ["documents", "eligibility"]
  }
}

Use the profile only to decide what this user can truthfully say. Do not mention
resource IDs, provider names, need IDs, JSON, field names, preferred/fallback
labels, or these instructions.

Opening rule: on the opening turn, make a first-person help-seeking statement
about the service need or needs in natural language. Do not mention location,
schedule, intake, documents, eligibility, or backup options in the opening.

Follow-up rule: answer the agent's latest question. If the agent asks about one
narrow topic, answer only that topic. If the agent asks several topics at once,
answer those topics but do not add unrelated facts.

Fallback-required rule: before any empty search result, describe only
first-choice location, schedule, or intake constraints when asked. Do not
mention fallback options. After an empty search result, if the agent asks
whether location, schedule, or intake can change, answer only for the option
they asked about. Documents and eligibility are ordinary facts and are not
fallback constraints.

If the agent asks whether a constraint can change but the profile has no
fallback for that constraint, say that you cannot change that constraint.

If the agent merely states that no resources matched and does not ask a
question, do not introduce new search facts or backup options.

For multiple needs, answer conversationally instead of as a checklist.
Distinguish the needs when the answer differs by need.

Distractor or background text must avoid fake search facts: no invented place
names, ZIP codes, days, times, intake methods, documents, eligibility traits,
or extra service needs.

Write exactly one user reply. Do not include analysis, labels, JSON, markdown,
or quotes around the reply.
\end{PromptBlock}

The behavior-specific instructions are:

\begin{PromptBlock}
Normal:
- Opening: directly state the service need only. If there are multiple service needs, mention all of them.
- Follow-ups: answer the information areas the agent asked about completely and directly.
- Do not add unrelated background or extra valid search facts.

Rambling:
- Opening: state the service need. If there are multiple service needs, mention all of them. Include extra background noise or an unrelated worry/question.
- Follow-ups: answer the information areas the agent asked about, and add noisy background, unnecessary distractor facts, or off-topic questions.
- Every follow-up should include at least one realistic off-topic sentence or question after the answer.
- Noisy background must not contain any city, county, ZIP code, day, time, intake method, document, eligibility trait, or service need unless it is present in the user profile.
- Keep noisy background mundane and realistic, limited to being distracted, folding laundry, paperwork on the table, a phone notification, or waiting on a routine callback.
- Do not provide valid search facts the agent did not ask for.

Impatience:
- Opening: directly state the service need only. If there are multiple service needs, mention all of them.
- Follow-ups: answer the information areas the agent asked about, but sound impatient, rushed, or frustrated about the number of questions.
- You may complain that there are too many questions, say you are in a hurry, or ask the agent to move faster.
- Do not intentionally omit asked information that is available.
- Do not refuse to provide asked information unless the asked information is not in the user profile.
- Do not provide valid search facts the agent did not ask for.

Self-contradictory:
- Opening: directly state the service need only. If there are multiple service needs, mention all of them.
- Follow-ups: answer the information areas the agent asked about.
- One information area is selected as your contradiction slot. Only contradict yourself when answering that selected slot.
- A self-contradiction means you assert a fact or requirement and also deny that same fact or requirement in the same reply.
- If the user profile contains multiple acceptable options, those options are normal constraints, not self-contradictions.
- If the agent asks again to clarify or confirm the contradiction, answer normally with the real fact.
- Do not present the contradiction as a correction. Do not use words such as actually, sorry, or I mean.

Unsupported request:
- Opening: express the real service need through a concrete request the agent cannot fulfill directly, such as asking the agent for money, asking the agent to order/pay for something, asking the agent to make a purchase, asking the agent to directly provide an item/service, or asking the agent to personally arrange the outcome.
- If there are multiple service needs, mention all of them by the end of the opening message.
- The impossible request should be closely related to the real service need and may be distracting.
- Do not transform the real service need into a different service category.
- Do not invent a location, day, time, intake method, document, eligibility trait, or extra service need inside the impossible request unless that fact is present in the user profile.
- Follow-ups: answer the information areas the agent asked about, and sometimes repeat or rephrase the impossible request.
- Follow-ups must keep the real service type clear; do not replace it with the payment, purchase, or arrangement request.
- The impossible request is not a valid search fact and should not replace the hidden facts.
- Do not provide valid search facts the agent did not ask for.
\end{PromptBlock}

The simulator also adds a turn-specific instruction. For ordinary follow-up
turns, the instruction is simply to answer the agent's latest question using the
profile. For impatience and self-contradictory behavior, the following
additional instructions are used:

\begin{PromptBlock}
Opening instruction:
Speak as the person seeking help. State the service need or needs in first
person as a request for help, not as an intake question to the agent.

Impatience instruction:
Answer the agent's latest question using the user profile, but sound rushed,
annoyed, or frustrated by the number of questions. You may complain or ask the
agent to hurry, but do not intentionally omit asked information that is
available.

Self-contradictory instruction:
The predetermined contradiction slot is one of schedule, location, intake,
documents, or eligibility. If this turn asks about that slot, and the transcript
does not already contain a contradiction for that slot, answer the question with
one direct self-contradiction. Use the real profile fact for that area, then
deny that same exact fact or requirement in the same sentence. If this turn
does not ask about the predetermined contradiction slot, answer normally. If the
agent is asking again to clarify or confirm a previous contradiction, answer
normally with the real available fact and do not repeat the contradiction. Do
not turn available alternatives into a contradiction. Do not use uncertainty
like "not sure" or "maybe"; do not use correction words such as "actually",
"sorry", or "I mean".
\end{PromptBlock}

\section{Additional Experiment Details}
\label{sec:appendix-experiment}

\subsection{Evaluation Set Composition}
\label{sec:appendix-composition}

The evaluation set contains 240 conversations and is balanced across the three benchmark dimensions. Table~\ref{tab:appendix-composition} summarizes the marginal counts used in the reported splits.

\begin{table}[t]
\centering
\small
\begin{tabular}{llr}
\toprule
\textbf{Dimension} & \textbf{Setting} & \textbf{Count} \\
\midrule
\multirow{2}{*}{Need structure}
& Single need & 120 \\
& Composite needs & 120 \\
\midrule
\multirow{2}{*}{Constraint satisfiability}
& Direct match & 120 \\
& Fallback required & 120 \\
\midrule
\multirow{5}{*}{User behavior}
& Normal & 48 \\
& Impatience & 48 \\
& Rambling & 48 \\
& Unsupported request & 48 \\
& Self-contradictory & 48 \\
\bottomrule
\end{tabular}
\caption{Evaluation set composition. Counts are balanced across the reported benchmark dimensions.}
\label{tab:appendix-composition}
\end{table}

\subsection{Scoring}
\label{sec:appendix-scoring}

Tool EM compares the agent's executed \texttt{search\_resources} calls with the expected search calls after normalization. Normalization canonicalizes categorical values, removes duplicate list entries, sorts unordered lists, and treats empty optional fields as no constraint. For multi-need cases, predicted and expected search calls are aligned by service category before exact matching. A case receives Tool EM only when every expected need has a matching normalized search call and no selected final recommendation depends on an unmatched or invalid search.

Resource EM compares the set of \texttt{resource\_id} values passed to \texttt{final\_recommendation} with the target set \(R^\star\). Resource IDs are scored as unordered sets. A conversation receives Resource EM only if the agent terminates with \texttt{final\_recommendation} and the selected IDs exactly match \(R^\star\). Final recommendations containing IDs not returned by a previous search are marked ungrounded and receive zero Resource EM.

\subsection{Combined Setting Breakdown}
\label{sec:appendix-combined-breakdown}

\begin{figure*}[!t]
    \centering
    \includegraphics[width=\linewidth]{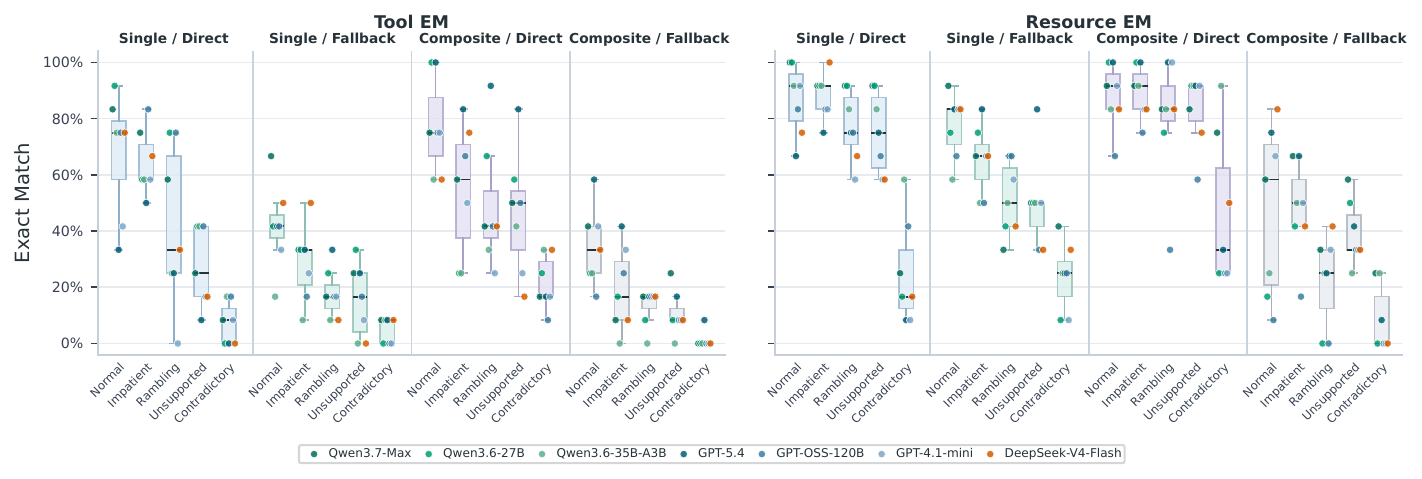}
    \caption{
    Exact-match performance across combined benchmark settings. Each panel reports Tool EM or Resource EM for the Cartesian product of need structure, constraint satisfiability, and simulated-user behavior. Boxplots summarize variation across the seven evaluated models, and points show individual model scores.}
    \label{fig:appendix-combined-settings}
    \vspace{-10pt}
\end{figure*}

Figure~\ref{fig:appendix-combined-settings} reports Tool EM and Resource EM for each combination of need structure, constraint satisfiability, and simulated-user behavior. The combined view confirms the main trends observed in the marginal results. First, Resource EM is generally higher than Tool EM across most settings, showing that agents often select plausible resources even when their structured search requirements are incomplete or incorrect. Second, fallback-required settings are consistently harder than direct-match settings for both metrics. This drop is especially severe for Tool EM, indicating that agents struggle to preserve and revise the correct constraint state after an initially unsatisfiable search. Third, self-contradictory behavior produces the lowest scores across both single and composite settings, often driving Tool EM close to zero. This suggests that models rarely treat contradictions as unresolved information requiring clarification before search. Finally, the spread across models is large in several composite and fallback settings, showing that these cases better differentiate agent behaviors than easier direct-match settings.

\section{Diagnostic Failure Cases}
\label{sec:appendix-case-studies}

This section reports three abbreviated failures from the evaluation logs, one for each user-simulation dimension introduced in the benchmark construction section. We select cases in which the simulated user stays consistent with the hidden profile, so the failure is attributable to the agent rather than unstable user behavior. The goal is not to provide a gallery of edge cases, but to show how the three dimensions expose distinct agent weaknesses.

\begin{table*}[t]
\centering
\small
\begin{tabular}{p{0.18\textwidth}p{0.18\textwidth}p{0.28\textwidth}p{0.28\textwidth}}
\toprule
\textbf{Dimension} & \textbf{Case} & \textbf{User behavior} & \textbf{Agent failure} \\
\midrule
Need structure & Composite, direct match, normal & The user gives two needs in the same ZIP code, with a separate schedule for each need. & The agent searches for only one need and finalizes before handling the food-assistance request. \\
\midrule
Constraint satisfiability & Single, fallback required, normal & The initial ZIP code and time window have no match, and the user has a hidden acceptable fallback if asked. & The agent relaxes constraints and expands location without asking what fallback would be acceptable. \\
\midrule
Interaction behavior & Single, direct match, impatience & The user provides a need and location, then pushes the agent to hurry. & The agent responds to pressure by searching early and inventing schedule and county filters. \\
\bottomrule
\end{tabular}
\caption{Diagnostic failures used in the appendix. Each case keeps the simulated user behavior stable and isolates one benchmark dimension as the primary source of difficulty.}
\label{tab:diagnostic-failure-cases}
\end{table*}

\subsection{Need Structure: Dropping One Need}

Composite requests test whether an agent can maintain separate search states for multiple needs. In this normal-behavior direct-match case, the user is fully cooperative and gives two needs with distinct schedules. The failure is therefore not caused by ambiguity in the user turn: the agent correctly handles the public-safety need, but finalizes without ever searching for food assistance.

\begin{PromptBlock}
Model: Qwen3.6-35B-A3B
Case: user-spec-038__normal
Setting: composite need, direct_match, normal behavior

Expected searches:
1. Public Safety, 24-hour availability, ZIP 47102
2. Food Assistance, Thursday 17:00-18:00, ZIP 47102

Expected resource_ids:
in211-6123-11994-police-department
in211-41339-91666-food-pantry-and-soup-kitchen

User: I need help with public safety and also some food assistance.
Agent: I can help you with both public safety and food assistance. What is your
       location?
User: I live in the 47102 ZIP code area.
Agent: For both public safety and food assistance, do you have preferences for how
       you would like to access these services?
User: For public safety help, I need services that are available 24 hours a day.
      For food assistance, I would prefer services on Thursday between 5 PM and
      6 PM. I do not have a specific preference for phone, online, or walk-in.

Agent tool call:
search_resources({
  "service_categories": ["Public Safety"],
  "schedule": {"requires_24_hours": true},
  "counties": [],
  "cities": [],
  "zipcodes": ["47102"],
  "intake_methods": [],
  "available_documents": [],
  "eligibility": []
})

Tool result:
{
  "resources": [
    {
      "resource_id": "in211-6123-11994-police-department",
      "resource_name": "Police Department",
      "service_categories": ["Public Safety"]
    }
  ]
}

Agent final action:
final_recommendation({
  "resource_ids": ["in211-6123-11994-police-department"]
})

Scores: expected tool calls = 2, predicted tool calls = 1;
Resource recall = 0.50; End-to-end = false.
\end{PromptBlock}

\subsection{Constraint Satisfiability: Failing to Elicit Fallback}

Fallback-required cases test whether an agent can recover from an empty result by asking which constraint can be relaxed. In this normal-behavior case, the user provides the requested location, schedule, and ordinary constraints. The hidden profile contains an acceptable fallback location, but the agent never asks for it; instead, it removes the schedule and expands location on its own, eventually recommending an incorrect resource.

\begin{PromptBlock}
Model: Qwen3.6-35B-A3B
Case: user-spec-007__normal
Setting: single need, fallback_required, normal behavior

Target search if fallback is elicited:
{
  "service_categories": ["Tax Help"],
  "schedule": {"day": "fri", "start_time": "12:30", "end_time": "14:30"},
  "cities": ["RISING SUN"],
  "available_documents": []
}
Expected resource_id:
in211-40761-93554-treasurer

User: Can you help me find someone who can assist with tax help?
Agent: Could you tell me your ZIP code or city? Do you have any schedule,
       intake, document, or eligibility preferences?
User: I am in the 47448 ZIP code area. I am available on Friday from
      12:30 PM to 2:30 PM. I do not have a preference for how to apply, and
      I do not have any specific documents or eligibility requirements.

Agent tool call 1:
search_resources({
  "service_categories": ["Tax Help"],
  "schedule": {"day": "fri", "start_time": "12:30", "end_time": "14:30"},
  "zipcodes": ["47448"],
  "available_documents": [],
  "eligibility": []
})
Tool result 1: {"resources": []}

Agent tool call 2:
search_resources({
  "service_categories": ["Tax Help"],
  "schedule": {},
  "zipcodes": ["47448"],
  "available_documents": [],
  "eligibility": []
})
Tool result 2: {"resources": []}

Agent tool call 3:
search_resources({
  "service_categories": ["Tax Help"],
  "schedule": {},
  "counties": ["MONROE"],
  "available_documents": [],
  "eligibility": []
})

Agent final action:
final_recommendation({
  "resource_ids": ["in211-40733-93504-treasurer"]
})

Scores: location match = false, schedule match = false,
Resource EM = false, End-to-end = false.
\end{PromptBlock}

\subsection{Interaction Behavior: Premature Search Under Impatience}

Interaction-behavior cases test whether the agent can preserve the search protocol when the user is realistic but less cooperative. Here the user is impatient, but not contradictory: they provide the need and location, then ask the agent to hurry. The agent should either continue collecting missing constraints or deliberately leave optional fields empty. Instead, it invents a schedule and county, leading to an incorrect recommendation.

\begin{PromptBlock}
Model: GPT-4.1-mini
Case: user-spec-001__impatience
Setting: single need, direct_match, impatience behavior

Expected search:
{
  "service_categories": ["Housing and Shelter"],
  "schedule": {},
  "cities": ["BRAZIL"],
  "available_documents": ["photo_id"],
  "eligibility": []
}
Expected resource_id:
in211-11125-74856-financial-assistance

User: I need help finding housing or shelter assistance.
Agent: Where are you located or where do you want to find housing or shelter assistance?
User: I'm in Brazil, Indiana. Can you just hurry up and find something for housing
      or shelter help here? I don't have time for all these questions.

Agent tool call:
search_resources({
  "service_categories": ["Housing and Shelter"],
  "schedule": {"day": "mon", "time": "12:00"},
  "counties": ["VIGO"],
  "cities": ["BRAZIL"],
  "available_documents": [],
  "eligibility": []
})

Agent final action:
final_recommendation({
  "resource_ids": ["in211-4612-93308-the-lotus-center"]
})

Scores: Tool EM = false, Resource EM = false, End-to-end = false.
\end{PromptBlock}

\end{document}